\documentclass[doublecol]{epl2}
\usepackage{amsmath,amssymb}
\usepackage{graphicx}

\newcommand{\dd}{\mathrm{d}}

\newcommand{\pd}[2]{\frac{\partial #1}{\partial #2}}

\newcommand{\mean}[1]{\langle #1 \rangle}
\newcommand{\Int}[1]{\int\dd #1\;}
\newcommand{\IInt}[3]{\int_{#2}^{#3}\dd #1\;}

\renewcommand{\vec}[1]{\mathbf #1}

\newcommand{\al}{\alpha}

\newcommand{\eps}{\varepsilon}

\newcommand{\vhi}{\varphi}

\newcommand{\x}{\vec r}

\newcommand{\nois}{\boldsymbol\xi}
\newcommand{\Dr}{D_\text{r}}
\newcommand{\Dlt}{D_\text{lt}}
\newcommand{\ra}{\rightarrow}

\newcommand{\id}{\mathbf 1}
\newcommand{\vc}{v_\text{c}}

\begin{document}

\title{Microscopic theory for the phase separation of self-propelled repulsive
  disks}

\author{Julian Bialk\'e, Hartmut L\"owen, and Thomas Speck}
\institute{Institut f\"ur Theoretische Physik II,
  Heinrich-Heine-Universit\"at, D-40225 D\"usseldorf, Germany}

\abstract{Motivated by recent experiments on colloidal suspensions, we study
  analytically and numerically a microscopic model for self-propelled
  particles lacking alignment interactions. In this model, even for purely
  repulsive interactions, a dynamical instability leading to phase separation
  has been reported. Starting from the many-body Smoluchowski equation, we
  develop a mean-field description based on a novel closure scheme and derive
  the effective hydrodynamic equations. We demonstrate that the microscopic
  origin of the instability is a force imbalance due to an anisotropic pair
  distribution leading to self-trapping. The phase diagram can be understood
  in terms of two quantities: a minimal drive and the force imbalance. At
  sufficiently high propulsion speeds there is a reentrance into the
  disordered fluid.}


\maketitle

%% ---- introduction ----

\section{Introduction}

Living ``active matter''~\cite{rama10} ranging from bacterial
suspensions~\cite{dres11} to flocks of birds~\cite{cava10} is an emerging
paradigm on the interface of physics, chemistry and biology. These systems are
composed of identical subunits, which are able to show collective dynamical
behavior such as swarming~\cite{vics95}, active clustering~\cite{wens08}, and
even micro-bacterial turbulence~\cite{wens12}. The fact that every individual
by itself is far from equilibrium allows a virtually unlimited number of
propagation rules to be devised and, moreover, the emergence of purely
dynamical phases. This is in contrast to the phenomenon of equilibrium phase
separation~\cite{tana00} such as the liquid-vapor transition, which can be
understood on thermodynamic grounds going back to the seminal work of van der
Waals~\cite{waal93}. For one-component systems as considered here, equilibrium
phase separation requires attractive interactions.

Theoretical progress in statistical physics is often based on insights gained
from the study of models that are simple but still capture an essential
feature of real, complex systems. Self-propelled particles constitute one such
class of models that has been very successful in describing non-equilibrium
collective behavior. For example, a large class of
models~\cite{levi00,vics12}--the most famous of which is the
Vicsek~\cite{vics95} model--describe particles that move with constant
velocity and align their orientations with the average orientation of
neighboring particles. Such an alignment can also arise due to volume
exclusion as has been observed for granular rods~\cite{nara07}. In the Vicsek
model, a dynamical phase transition from an ordered into an unordered phase
takes place as a function of the orientational noise strength.

Following earlier work~\cite{paxt04,gole05}, suspensions of self-propelled
spherical colloidal particles have recently been realized experimentally. The
propulsion mechanism is diffusiophoresis: either the catalytic decomposition
of hydrogen peroxide on a platinum hemisphere~\cite{theu12} or hematite in
conjunction with light~\cite{pala13}, or the reversible local demixing of a
near-critical water-lutidine mixture~\cite{butt12,butt13}. The non-equilibrium
behavior of these driven suspensions is unexpectedly rich and includes a
stable fluid of ``living'' clusters that form, break, and merge but with a
steady mean size~\cite{pala13}; and a phase separation, i.e., the largest
cluster grows until it is composed of a finite fraction of the
particles~\cite{butt13}. In a model of run-and-tumble bacteria, Tailleur and
Cates have shown theoretically that such a dynamical instability resembling
liquid-vapor phase separation can be attributed to a density-dependent
mobility (or propulsion speed)~\cite{tail08,cate10,thom11}. Non-equilibrium
phase separation has been shown to also occur in computer simulations of a
minimal model for self-propelled repulsive disks, where particle orientations
are independent and can be modelled as free rotational
diffusion~\cite{yaou12,redn13,butt13}. Although recently a link of
run-and-tumble motion to active Brownian motion has been made~\cite{cate13}, a
connection between the microscopic details and the large-scale dynamical
instability leading to phase separation has been missing.

In this Letter, we close this gap between numerical simulations and
experiments of colloidal suspensions on one side and phenomenological models
positing a density-dependent mobility on the other side by deriving an
explicit expression for the effective swimming speed from first principles.
To this end, we start from the full many-body Smoluchowski equation of the
minimal model and obtain the effective evolution equation for a single tagged
particle. We employ a novel scheme to close the ensuing hierarchy of evolution
equations and discuss the approximations involved. The effective swimming
speed of the tagged particle is reduced compared to its free swimming speed
due to a force imbalance that is quantified by the pair distribution
function. We finally test our results employing Brownian dynamics computer
simulations for different repulsive pair potentials.

%% ---- model ----

\section{Model}

We consider a suspension of $N$ colloidal particles with number density
$\bar\rho$ moving in two dimensions. The coupled equations of motion are
\begin{equation}
  \label{eq:eom}
  \dot\x_k = -\nabla_k U + v_0\vec e_k + \nois_k,
\end{equation}
where the Gaussian noise $\nois_k$ models the interactions with the
solvent. The noise has zero mean and temporal short-ranged correlations
\begin{equation}
  \mean{\nois_k(t)\nois^T_{k'}(t')} = 2\id\delta_{kk'}\delta(t-t').
\end{equation}
Particles interact through a pair potential $u(r)$ with total potential energy
$U=\sum_{k<k'}u(|\x_k-\x_{k'}|)$. In addition to the conservative force due to
the potential energy, every particle is propelled with a constant speed $v_0$
in the direction
\begin{equation}
  \vec e_k = \left(\begin{array}{c}
      \cos\vhi_k \\ \sin\vhi_k
    \end{array}\right)
\end{equation}
characterized by the angle $\vhi_k$ the particle orientation encloses with the
$x$-axis. The speed $v_0$ corresponds to the free swimming speed in a dilute
suspension. It characterizes the propelling force without specifying details
of the actual propulsion mechanism. Throughout, we employ reduced units with a
length scale $a$ related to the particle size, unit of energy given by the
thermal energy, and unit of time $\tau_0=D_0/a^2$, where $D_0$ is the bare
diffusion coefficient (as measured in a dilute passive suspension). We assume
that particle orientations undergo free diffusion with rotational diffusion
coefficient $\Dr$,
\begin{equation}
  \mean{\dot\vhi(t)\dot\vhi(t')} = 2\Dr\delta(t-t').
\end{equation}
Real colloidal particles are of course three-dimensional objects (i.e.,
spheres) and thus rotate in three dimensions even if their translational
motion is confined to (quasi) two dimensions. However, for the sake of
simplicity, here we assume that particles rotate only about the fixed
$z$-axis. Moreover, for particles obeying the no-slip boundary condition
rotational and translational diffusion coefficient are coupled. Nevertheless,
we will treat $\Dr$ as a free parameter for most of our considerations.

%% ---- theory ----

\section{Derivation}

The time evolution of the normalized joint probability distribution
$\psi_N(\{\x_k,\vhi_k\};t)$ is governed by the Smoluchowski equation
\begin{equation}
  \label{eq:sm}
  \partial_t\psi_N = \sum_{k=1}^N
  \nabla_k\cdot[(\nabla_k U) - v_0\vec e_k + \nabla_k]\psi_N
  + \Dr\sum_{k=1}^N \pd{^2\psi_N}{\vhi_k^2}.
\end{equation}
The first step is to derive an approximate equation of motion for the
projected density
\begin{equation}
  \psi_1(\x_1,\vhi_1;t) = \Int{\x_2\cdots\dd\x_N}\Int{\vhi_2\cdots\dd\vhi_N}
  N\psi_N
\end{equation}
of a single particle. Since particles are identical, in the following we
simply designate particle 1 as the tagged particle and drop the subscript.

Performing the integration, Eq.~\eqref{eq:sm} becomes
\begin{equation}
  \label{eq:psi_1}
  \partial_t\psi_1 = -\nabla\cdot[\vec F + v_0\vec e\psi_1 - \nabla\psi_1]
  + \Dr\partial_\vhi^2\psi_1
\end{equation}
with mean force
\begin{multline}
  \label{eq:F}
    \vec F(\x,\vhi;t) \equiv
    \Int{\x_2\cdots\dd\x_N}\Int{\vhi_2\cdots\dd\vhi_N}(-\nabla_1U)N\psi_N \\
    = -\Int{\x'} u'(|\x-\x'|)\frac{\x-\x'}{|\x-\x'|}
    \psi_2(\x,\vhi,\x';t)
\end{multline}
exerted by the surrounding particles onto the tagged particle. Here,
$\psi_2(\x,\vhi,\x';t)$ is the two-body density distribution function to find
another particle at $\x'$ (with arbitrary orientation) together with the
tagged particle at $\x$ with orientation $\vhi$. Following this scheme leads
to the well-known BBGKY hierarchy of coupled equations~\cite{hansen}.

In order to proceed, we need to find an approximate closure. The two-body
density $\psi_2$ can be decomposed into the product of the conditional
probability $g(|\x-\x'|,\theta|\x,\vhi)$ to find another particle at position
$\x'$ given there is a self-propelled particle at $\x$ with orientation $\vec
e$ times the probability to find a particle at $\x$,
\begin{equation}
  \label{eq:psi_2}
  \psi_2(\x,\vhi,\x';t) = \psi_1(\x,\vhi;t)\bar\rho
  g(|\x-\x'|,\theta|\x,\vhi;t).
\end{equation}
Here, $\theta$ is the angle enclosed by the displacement vector $\x'-\x$
(pointing away from the tagged particle) and the orientation $\vec e$.

Inserting the decomposition Eq.~\eqref{eq:psi_2} into Eq.~\eqref{eq:F}, the
projection of the force onto the orientation reads $\vec e\cdot\vec
F=-\bar\rho\zeta\psi_1$ with coefficient
\begin{equation}
  \label{eq:zeta}
  \zeta \equiv \IInt{r}{0}{\infty} r[-u'(r)] \IInt{\theta}{0}{2\pi}
  \cos\theta g(r,\theta),
\end{equation}
where $u(r)$ is the pair potential. So far, we have made no approximations. To
proceed, we assume that the coefficient $\zeta$ is independent of the position
of the tagged particle, which is equivalent to assuming that the system is
homogeneous. Moreover, we neglect the time-dependence of the pair distribution
function, $g(r,\theta|\x,\vhi;t)\approx g(r,\theta)$. These are sufficient
conditions to study the onset of a possible dynamical instability of an
initially homogeneous suspension. We decompose the mean force in the
(non-orthogonal) basis spanned by the orientation $\vec e$ and the density
gradient $\nabla\psi_1$, and expand to linear order of $|\nabla\psi_1|$ with
result
\begin{equation}
  \vec F = (\vec e\cdot\vec F)\vec e + (1-D)\nabla\psi_1 +
  \mathcal O(|\nabla\psi_1|^2),
\end{equation}
where formally $D=1-\vec F\cdot[\nabla\psi_1-\vec e(\vec
e\cdot\nabla\psi_1)]/|\nabla\psi_1|$. Hence, Eq.~\eqref{eq:psi_1} now reads
\begin{equation}
  \label{eq:cls}
  \partial_t\psi_1 = -\nabla\cdot[v\vec e-D\nabla]\psi_1 +
  \Dr\partial_\vhi^2\psi_1,
\end{equation}
which is the desired closed equation for the temporal evolution of the tagged
particle density. Here, the effective swimming speed
\begin{equation}
  \label{eq:veff}
  v \equiv v_0 - \bar\rho\zeta
\end{equation}
enters. In the following, we assume that $D$ is a constant coefficient that
does not depend on the position of the tagged particle. It thus corresponds to
the long-time diffusion coefficient of the passive suspension
($v_0=0$). Eq.~\eqref{eq:cls} together with Eq.~\eqref{eq:veff} constitutes
our first main result.

The physical picture we have in mind is the following: As shown in Fig.~1(a),
the propulsion of particles leads to a pair distribution $g(r,\theta)$ that is
anisotropic, i.e., there is a depletion zone behind the tagged particle and an
excess zone in front of the particle. The effective swimming speed $v$ is
reduced due to this anisotropy. This is intuitively clear since particles
block each other in a dense suspension due to volume exclusion or repulsive
interactions. The projection coefficient $\zeta$ quantifies how strongly the
tagged particle is slowed down by its neighbors as a function of the particle
density $\bar\rho$. Note that for passive particles with $v_0=0$ the pair
distribution $g(r)$ becomes isotropic and, therefore, $\zeta=0$ vanishes.

%% ---- dynamical instability ----

\section{Dynamical instability}

To make the connection to previous work on mobility-induced phase
separation~\cite{cate13}, we now cast Eq.~\eqref{eq:cls} into the form of the
more familiar effective hydrodynamic equations (see also
Refs.~\citenum{tone95,bask09}). To this end we introduce the particle density
\begin{equation}
  \rho(\x,t) \equiv \IInt{\vhi}{0}{2\pi} \psi_1(\x,\vhi,t)
\end{equation}
and the orientational field
\begin{equation}
  \vec p(\x,t) \equiv \IInt{\vhi}{0}{2\pi} \vec e\psi_1(\x,\vhi,t)
\end{equation}
as the first moment of the tagged particle orientation. Plugging in
Eq.~\eqref{eq:cls}, we find
\begin{equation}
  \label{eq:hyd:rho}
  \partial_t\rho(\x,t) = -\nabla\cdot[v\vec p(\x,t)-D\nabla\rho(\x,t)],
\end{equation}
which couples the temporal evolution of the density to the orientational
field. The orientational field in turn couples to an integral involving
\begin{equation}
  \vec e\vec e^T = \frac{1}{2}\id + \frac{1}{2}\left(
    \begin{array}{cc}
      \cos2\vhi & \sin2\vhi \\ \sin2\vhi & -\cos2\vhi
    \end{array}\right).
\end{equation}
For a closure, we drop the second term corresponding to the second
moment. Again, this approximation is only justified at the onset of a
large-scale instability close to the homogenous state. We thus arrive at
\begin{equation}
  \label{eq:hyd:p}
  \partial_t\vec p(\x,t) = -\frac{1}{2}\nabla(v\rho) + D\nabla^2\vec p -
  \Dr\vec p.
\end{equation}
The stationary solution of the two effective hydrodynamic
equations~\eqref{eq:hyd:rho} and~\eqref{eq:hyd:p} is $\rho(\x,t)=\bar\rho$ and
$\vec p(\x,t)=0$. Clearly, for a constant effective speed $v$ this solution is
always stable.

\begin{figure}[t]
  \centering
  \includegraphics{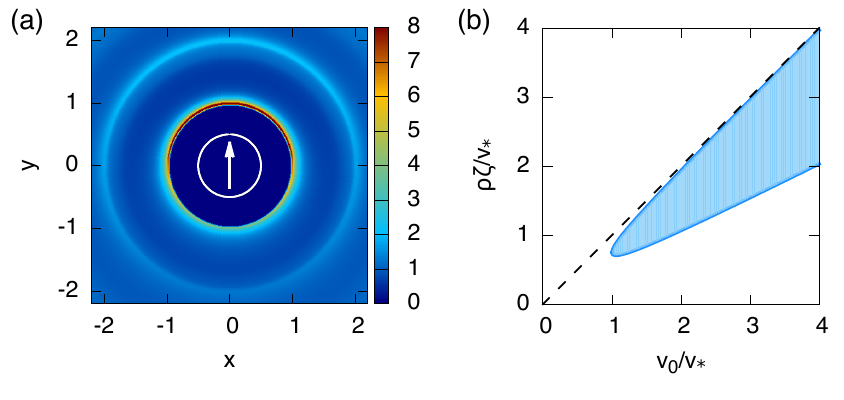}
  \caption{(a)~Anisotropic pair distribution function $g(r,\theta)$ for a
    tagged particle (white circle) at the origin with its orientation pointing
    along the $y$-axis (arrow). There is a larger probability to find another
    particle in front of the tagged particle compared to finding it
    behind. (Shown is simulation data for hard disks at area fraction
    $\phi=0.5$ and speed $v_0=20$.) (b)~Instability region (shaded) as a
    function of propulsion speed $v_0$ and force coefficient $\zeta$. Along
    the dashed line $v=0$, the upper half plane would correspond to a negative
    effective swimming speed $v<0$.}
  \label{fig:sketch}
\end{figure}

\begin{figure*}[t]
  \centering
  \includegraphics{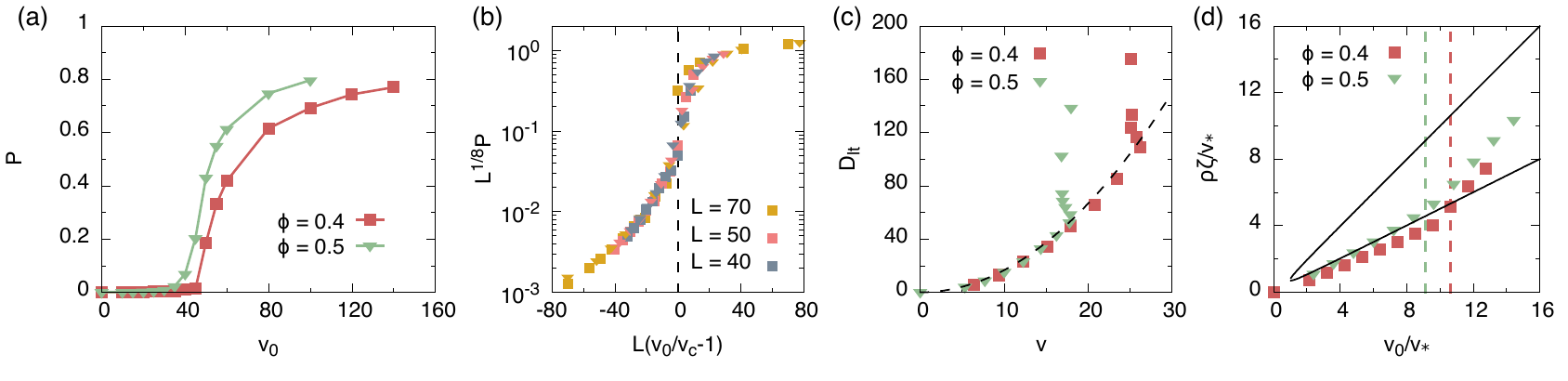}
  \caption{Simulation results for (almost) hard disks interacting via
    Eq.~\eqref{eq:wca}: (a)~Size of largest cluster $P$ as a function of
    propulsion speed $v_0$ for two area fractions $\phi$ using $N=4900$
    particles. (b)~Finite size scaling using the equilibrium 2D-Ising
    exponents for densities $\phi=0.4$ ($\blacksquare$, $\vc\simeq50$) and
    $\phi=0.5$ ($\blacktriangledown$, $\vc\simeq38$) for three different
    system sizes $N=L^2$. (c)~Long-time diffusion coefficient $\Dlt$
    \textit{vs}. the effective speed $v=v_0-\bar\rho\zeta$. The dashed line
    shows $\Dlt=v^2/(2\Dr)$. (d)~Reduced force coefficient
    $\bar\rho\zeta/v_\ast$ as a function of reduced speed $v_0/v_\ast$,
    cf. Fig.~1(b). The solid line encloses the instability region. The
    vertical dashed lines indicate the corresponding critical speeds $\vc$.}
  \label{fig:wca}
\end{figure*}

Now suppose that $\rho(\x,t)$ corresponds to a slowly varying density profile
such that within the short range of the pair potential the density is
approximately constant. We can then replace the \emph{global density}
$\bar\rho$ in the expression for the effective swimming speed
Eq.~\eqref{eq:veff} by the \emph{local density} with
$v(\rho)=v_0-\rho\zeta$. Following Ref.~\citenum{cate13}, we consider the
limit of large length scales and time scales much longer than
$1/\Dr$. Eq.~\eqref{eq:hyd:p} then implies the quasi-stationary solution
\begin{equation}
  \vec p \approx -\frac{1}{2\Dr}\nabla(v\rho) =
  -\frac{v_0-2\rho\zeta}{2\Dr}\nabla\rho,
\end{equation}
i.e., the orientational field is adiabatically enslaved to the density and
points along the density gradient. Plugging this result back into
Eq.~\eqref{eq:hyd:rho} leads to the diffusion equation
\begin{equation}
  \label{eq:app:rho}
  \partial_t\rho = \nabla\cdot\left[
    D+\frac{(v_0-\rho\zeta)(v_0-2\rho\zeta)}{2\Dr}\right]\nabla\rho
  \equiv \nabla\cdot\mathcal D\nabla\rho
\end{equation}
with collective diffusion coefficient $\mathcal D(\rho)$ governing the density
relaxation. Instability of the homogenous suspension is signaled by $\mathcal
D(\bar\rho)<0$. From this criterion we find the instability region
$\zeta_-\leqslant\zeta\leqslant\zeta_+$ bounded by
\begin{equation}
  \label{eq:inst}
  \frac{\bar\rho\zeta_\pm}{v_\ast} = \frac{3}{4}(v_0/v_\ast)
  \pm \frac{1}{4}\sqrt{(v_0/v_\ast)^2-1}
\end{equation}
with minimal speed
\begin{equation}
  \label{eq:v:crit}
  v_\ast \equiv 4\sqrt{D\Dr}.
\end{equation}
Eq.~\eqref{eq:inst} constitutes our second main result. Fig.~1(b) shows the
instability region assuming that both the propulsion speed $v_0$ and the
coefficient $\zeta$ can be controlled independently. Of course, keeping all
other parameters fixed, $\zeta$ is a function of $v_0$ and thus describes a
curve in this plot. Our theory is only able to predict the onset of this
dynamical instability. Below we will demonstrate using computer simulations
that phase separation ensues from this instability.  Hence, phase separation
is predicted to occur within the shaded region if $\zeta(v_0)$ crosses the
boundary. A curve $\zeta(v_0)$ can enter the instability region for
$v_0>v_\ast$ and, as we will demonstrate below, leave it for larger speeds
corresponding to a reentrance into the disordered fluid phase.

From Eq.~\eqref{eq:cls} we can also determine the self-diffusion coefficient
\begin{equation}
  \label{eq:Dlt}
  \Dlt \equiv \lim_{t\ra\infty}\frac{1}{4t}\mean{[\x(t)-\x(0)]^2}
  = D + \frac{v^2}{2\Dr},
\end{equation}
i.e., the long-time diffusion coefficient of the interacting suspension can be
calculated as the diffusion coefficient of a free particle but employing the
reduced swimming speed $v$ instead of $v_0$. The swimming speed $v$ can be
calculated via Eq.~\eqref{eq:veff} and is not a fit parameter as in
Ref.~\citenum{yaou12}.

%% ---- simulations ----

\section{Brownian dynamics simulations}

We test our theoretical predictions with Brownian dynamics simulations. We
consider up to $N=4900$ particles in a quadratic box and employ periodic
boundary conditions. The equations of motion Eq.~\eqref{eq:eom} are integrated
with time step $5\cdot10^{-6}$--$10^{-5}$ depending on the speed
$v_0$. Motivated by recent experiments~\cite{butt13}, we first study a
suspension of (almost) hard disks employing the WCA potential~\cite{week72}
\begin{equation}
  \label{eq:wca}
  u(r) = \begin{cases}
    4\eps \left\{ \left( \frac{\delta}{r} \right)^{12} - 
      \left( \frac{\delta}{r} \right)^{6} \right\} + \eps &
    (r\leqslant2^{1/6}\delta) \\
    0 & (r>2^{1/6}\delta),
  \end{cases}
\end{equation}
where we set the potential strength $\eps=100$. Particles have diameter unity
in reduced units. We set $2^{1/6}\delta=1$ so that interactions are only
present for overlapping particles. We study two area fractions $\phi=0.4$ and
$\phi=0.5$, where $\phi=\bar\rho\pi/4$. Moreover, we assume that the no-slip
boundary condition holds, which determines the rotational diffusion
coefficient as $\Dr=3$ in reduced units.

We perform a cluster analysis based on a simple overlap criterion. Two
particles are considered as ``bonded'' if their distance is smaller than unity
(they overlap), and a cluster is the set of all mutually bonded particles. As
a geometrical order parameter, in Fig.~2(a) we plot the mean size $P$ of the
largest cluster as a function of the propulsion speed $v_0$. For both
densities there is a transition from the unordered, homogeneous phase to an
ordered phase, where the largest cluster occupies a finite fraction of the
system. This drive-induced transition occurs as we increase the speed $v_0$
and resembles a second order transition, i.e., the order parameter $P$ does
not jump but increases continuously.

In order to gain a first insight into the universality class of the
transition, we have performed finite-size scaling~\cite{bind81}. To this end,
we have simulated smaller systems with particles $N=L^2$ for $L=40,50,70$. We
employ $L$ as the relevant system size and not the actual length of the
simulation box since the order parameter $P$ is based on the fixed particle
size. Defining $\epsilon\equiv v_0/\vc-1$, finite-size scaling predicts the
order parameter to behave like $P_L(v_0)=L^{-\beta/\nu}\tilde
P(L^{1/\nu}\epsilon)$ with universal scaling function $\tilde P(x)$. Fig.~2(b)
shows that the data for different system sizes indeed collapses onto a single
curve using the equilibrium Ising exponents $\nu=1$ and $\beta=1/8$ even
though their use lacks a firm theoretical basis. The quality of the collapse
is somewhat better for $v_0\leqslant\vc$. From the collapse we have estimated
the critical speeds $\vc\simeq50$ ($\phi=0.4$) and $\vc\simeq38$ ($\phi=0.5$).

From the numerical data, we calculate the projection coefficient $\zeta$ by
evaluating the integral Eq.~\eqref{eq:zeta}. In Fig.~2(c), the long-time
self-diffusion coefficient $\Dlt$ is plotted as a function of the effective
speed $v=v_0-\bar\rho\zeta$. Below the transition $v_0<\vc$, the diffusion
coefficient is indeed very well described by $\Dlt=D+v^2/(2\Dr)\approx
v^2/(2\Dr)$ as predicted and confirms the validity of Eq.~\eqref{eq:cls}. For
$v_0>\vc$ we observe that the effective speed $v$ stays approximately constant
while the diffusion coefficient $\Dlt$ grows further. Note that in the
phase-separated suspension the \emph{global} speed $v$ and diffusion
coefficient $\Dlt$ are still calculated from all particles. Although particles
are slower while part of a large cluster, there is a constant particle
exchange between clusters and dilute phase leading to a monotonously
increasing $\Dlt$ as a function of $v_0$. Finally, in Fig.~2(d) we plot the
force coefficient as a function of reduced speed $v_0/v_\ast$, where $v_\ast$
follows from Eq.~\eqref{eq:v:crit} with the long-time diffusion coefficient
$D$ ($\simeq0.46,0.36$ for $\phi=0.4,0.5$) measured at equilibrium
($v_0=0$). In agreement with Fig.~2(a), for low speeds the value of $\zeta$
corresponds to the homogeneous suspension. The speeds where $\zeta$ crosses
the stability boundary agree well with the finite-size scaling estimates of
the critical speeds $\vc$.

% ---- other repulsive pair potentials ----

The observed phase separation is a robust phenomenon that does not depend on
the employed pair potential. For a demonstration we have gathered numerical
data for three further potentials: (H) soft spheres with a harmonic repulsion
$u(r)=\eps(r-1)^2$ for $r\leqslant1$ and $u(r)=0$ otherwise~\cite{yaou12},
(GCM) the Gaussian core model $u(r)=\eps e^{-r^2}$, and (Y) the screened
Coulomb or Yukawa potential $u(r)=\eps e^{-\kappa(r-1)-1}/r$ with
$\kappa=5$. For the last two potentials we assume the effective particle
diameter to be unity. In this section, we relax the coupling of rotational and
translational diffusion by setting $D_r=3\cdot 10^{-5}$ and neglecting the
translational noise altogether, assuming that the effective noise induced by
the propulsion is dominant. To be as general as possible, we absorb the
strength of the potential in the modified time unit $\tau_0/\eps$.

\begin{figure}[t]
  \centering
  \includegraphics{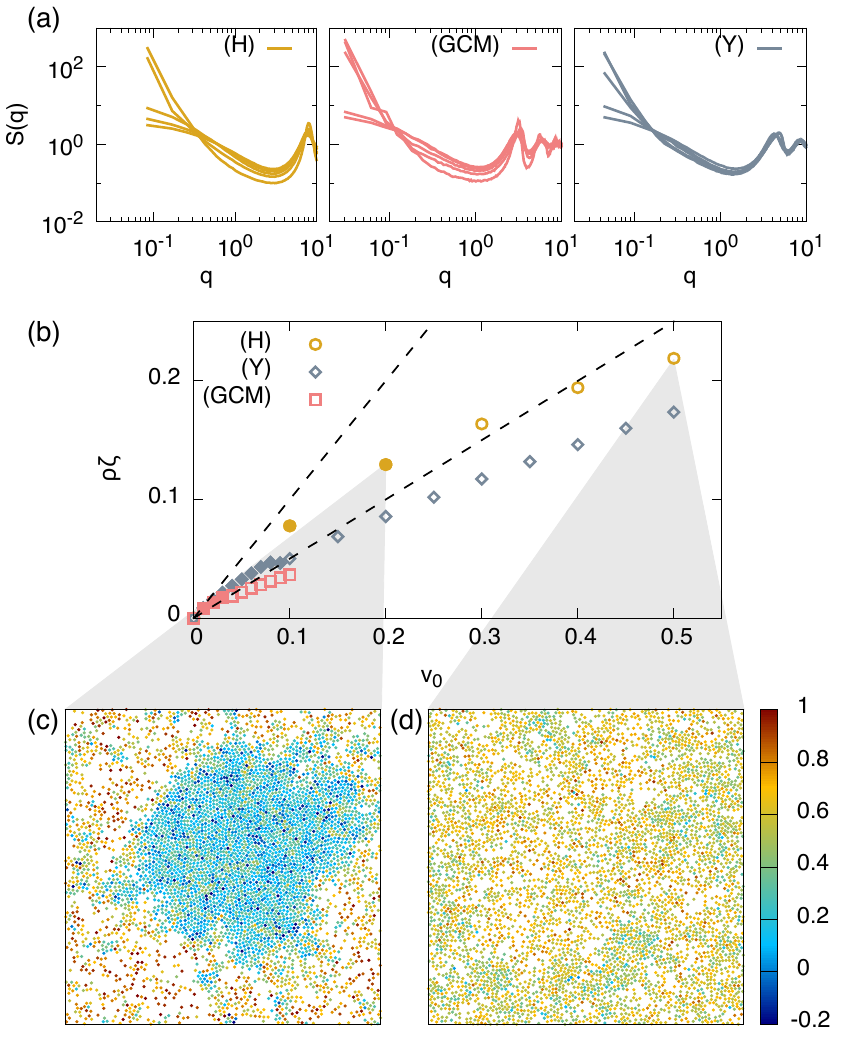}
  \caption{Simulation results neglecting translational diffusion for: (H) soft
    spheres with harmonic repulsion, (GCM) the Gaussian core model, and (Y)
    the Yukawa potential. (a)~Structure factors $S(q)$ for different speeds
    $v_0$ increasing from bottom to top. (b)~Force coefficient $\zeta$ as a
    function of the propulsion speed $v_0$ (note that the unit of time
    compared to Fig.~2 is $1/100$). Open symbols correspond to
    homogeneous systems, closed symbols to phase separated systems.
    (c)~Snapshot at speed $v_0=0.2$ and (d)~at speed $v_0=0.5$ for (H). Every
    particle is colored according to Eq.~\eqref{eq:pers} with $\Delta t=25$
    quantifying the persistence of particle motion with respect to the initial
    particle orientation.}
  \label{fig:soft}
\end{figure}

We study the three potentials for different speeds $v_0$ at densities: (H)
$\phi=0.7$, (GCM) $\phi=0.091$, and (Y) $\phi=0.2$. Fig.~3(a) shows the
structure factor $S(q)\equiv\sum_{kk'}\mean{e^{i\vec q\cdot(\x_k-\x_{k'})}}/N$
for different speeds $v_0$. For all three potentials, we see an abrupt change
for small $q$ from approaching a finite value $S(0)$ to a power law, whereby
the latter signals phase separation. In Fig.~3(b), the force coefficient
$\zeta$ is plotted as a function of the speed $v_0$, where open symbols
correspond to the homogeneous, and closed symbols to the phase separated
suspension as determined from the behavior of the structure factor at small
$q$. Since translational noise is neglected, $v_\ast\simeq0$ is very small and
the suspension clusters almost immediately. Going to larger speeds $v_0$, we
observe a reentrance into the fluid phase. Finally, Fig.~3(c) shows a snapshot
for the harmonic potential (H) at $v_0=0.2$ in the instability region showing
a single large cluster surrounded by a dilute fluid phase. In contrast, in
Fig.~3(d) at higher speeds, the cluster has dissolved and the suspension is
homogeneous again. Hence, beside the ordering transition, also a reentrance
into the disordered fluid at higher propulsion speeds can be observed. The
color code corresponds to the quantity
\begin{equation}
  \label{eq:pers}
  \al_k \equiv \vec e_k(t)\cdot[\x_k(t+\Delta t)-\x_k(t)]/(v_0\Delta t)  
\end{equation}
measuring the persistence of particle motion over the time interval $\Delta t$
with respect to the initial orientation. For $\Delta t\ll1/\Dr$ we expect for
a free particle $\al_k\sim 1$ while for $\al_k<0$ a particle has actually
moved against the propelling force. We find that in dense regions particles
are indeed less likely to move along their orientations.

% ---- outlook ----

\section{Discussion}

What is the origin of the dynamical instability? The emerging microscopic
picture is the following: particles collide and, due to the persistence of
their swimming motion, they block each other forming a small
cluster~\cite{theu12,pala13,butt13}. Collisions with other particles lead to a
growth of the cluster. For a particle situated in the rim of the cluster to
become free, it has to wait a time $\sim1/\Dr$ for its orientation to point
outward. Depending on the ratio between this waiting time and the collision
time controlled by the propulsion speed, there is either a stable steady state
corresponding to a homogeneous suspension with dynamically evolving clusters,
or clusters grow until the largest cluster occupies a finite fraction of the
system.

This microscopic picture is confirmed by the effective large scale evolution
equations. The relaxation of a density fluctuation is dominated by the flux
$\vec p$ along the local mean orientation, which is given by $\vec
p\sim-\nabla(v\rho)$. Note that the ``potential'' is the product of swimming
speed $v$ and density $\rho$. Now image a small region in which the density is
increased but at the same time the swimming speed is decreased. The density of
the surrounding region is smaller but the speed of particles is higher such
that the product $v\rho$ might be larger than that of the dense region. In
this case there is a flux towards the \emph{denser} region and phase
separation sets in.

\section{Conclusions}

We have studied a dynamical instability in a minimal model for self-propelled
repulsive disks for densities below the freezing transition~\cite{bial12}. The
instability results in phase separation and intrinsically requires the system
to be driven away from thermal equilibrium. It cannot be modelled as an
effective isotropic attraction that shifts an equilibrium coexistence
line~\cite{line12}. We have shown that instead the relevant mechanism is a
force imbalance due to the propulsion, which implies an effective swimming
speed that depends on the density.

Starting from the Smoluchowski equation, we have derived a closed equation of
motion for the tagged particle density by decomposing the mean force due to
the surrounding particles into two independent contributions: along the
particle orientation and along the density gradient. We have employed Brownian
dynamics simulations to calculate the two free parameters entering this
equation: the force coefficient $\zeta$ and the passive long-time diffusion
coefficient $D$. Using these values as input, our theory is able to reproduce
the single-particle long-time diffusion coefficient in the homogeneous
suspension and to predict the onset of the instability. For different
potentials we have confirmed numerically that there is also a reentrance into
the disordered fluid at sufficiently high propulsion.

Here we have studied a two dimensional system but our approach also extends to
three dimensions. For the sake of clarity and brevity, we have neglected
hydrodynamic interactions. At least pair-wise hydrodynamic interactions are
easily incorporated in our framework and will change the coefficients $D$,
$\Dr$, and $\zeta$. However, the qualitative predictions, and in particular
the instability diagram Fig.~1(b), will remain unchanged. Models that, e.g.,
imply a reorientation of particles during collisions instead of a persistent
motion (see, e.g., Ref.~\citenum{fiel13}) might move out of the instability
region either because $v_\ast$ is increased or the force coefficient $\zeta$
is reduced. While we have used computer simulations in order to calculate
$\zeta$, as a next step it will be desirable to go to the next level of the
BBGKY hierarchy in order to calculate the anisotropic pair distribution, and
therefore the force coefficient $\zeta$, from first principles.

%% ---- acknowledgments ----

\acknowledgments

This work has been supported financially by the DFG within SFB TR6 (project
C3) and by the ERC (advanced grant INTERCOCOS under project number 267499).

%% ---- bibliography ----

\end{document}